\documentclass[prd,preprint,tightenlines,floatfix,showpacs,preprintnumbers,nofootinbib,eqsecnum]{revtex4}

 \usepackage[dvips,final]{graphicx}
  \usepackage{amssymb}
   \usepackage{amsmath}
    \usepackage{amsfonts}
     \usepackage{epsfig}
      \usepackage{bm} 

\usepackage{slashed}      
      
\usepackage[colorlinks = true,
linkcolor = blue,
urlcolor  = blue,
citecolor = blue,
anchorcolor = blue]{hyperref}
      
\DeclareMathOperator*{\SumInt}{%
	\mathchoice%
	{\ooalign{$\displaystyle\sum$\cr\hidewidth$\displaystyle\int$\hidewidth\cr}}
	{\ooalign{\raisebox{.14\height}{\scalebox{.7}{$\textstyle\sum$}}\cr\hidewidth$\textstyle\int$\hidewidth\cr}}
	{\ooalign{\raisebox{.2\height}{\scalebox{.6}{$\scriptstyle\sum$}}\cr$\scriptstyle\int$\cr}}
	{\ooalign{\raisebox{.2\height}{\scalebox{.6}{$\scriptstyle\sum$}}\cr$\scriptstyle\int$\cr}}
}      
\begin{document}

\title{Chiral Vortical Effect in Extended Rarita-Schwinger Field Theory and Chiral Anomaly}

\author{G. Yu. Prokhorov}
\email{prokhorov@theor.jinr.ru}
\affiliation{Joint Institute for Nuclear Research, Joliot-Curie str. 6, Dubna 141980, Russia}
\affiliation{Institute of Theoretical and Experimental Physics, NRC Kurchatov Institute, 
B. Cheremushkinskaya 25, Moscow 117218, Russia}
\author{O. V. Teryaev}
\email{teryaev@jinr.ru}
\affiliation{Joint Institute for Nuclear Research, Joliot-Curie str. 6, Dubna 141980, Russia}
\affiliation{Institute of Theoretical and Experimental Physics, NRC Kurchatov Institute, 
	B. Cheremushkinskaya 25, Moscow 117218, Russia}
\author{V. I. Zakharov}
\email{vzakharov@itep.ru}
\affiliation{Institute of Theoretical and Experimental Physics, NRC Kurchatov Institute, 
B. Cheremushkinskaya 25, Moscow 117218, Russia}
\affiliation{Pacific Quantum Center, 
Far Eastern Federal University, 10 Ajax Bay, Russky Island, Vladivostok 690950, Russia\vspace{0.8 cm}}

\begin{abstract}
\vspace{0.8 cm}
We consider the theory of Rarita-Schwinger field interacting with a field with spin 1/2, in the case of finite temperature, chemical potential and vorticity, and calculate the chiral vortical effect for spin 3/2. We have clearly demonstrated the role of interaction with the spin 1/2 field, the contribution of the terms with which to CVE is 6. Since the contribution from the Rarita-Schwinger field is -1, the overall coefficient in CVE is 6-1=5, which corresponds to the recent prediction of a gauge chiral anomaly for spin 3/2.
The obtained values for the coefficients $ \mu^2 $ and $ T^2 $ are proportional to each other, but not proportional to the spin, which indicates a possible new universality between the temperature-related and the chemical potential-related vortical effects. The results obtained allow us to speculate about the relationship between the gauge and gravitational chiral anomalies.
\end{abstract}

\maketitle

\section{Introduction}
\label{sec:intro}

The Rarita-Schwinger spin 3/2 theory is an essential element of  supergravity theories \cite{Strathdee:1986jr} and grand unification models \cite{Adler:2014pga}, in which it is used for anomaly cancellation. Rarita-Schwinger fields are also used to describe hadronic resonances \cite{Pascalutsa:1999zz} and have applications in solid state physics when describing Rarita-Schwinger-Weyl semimetals \cite{Boettcher:2019dtq}.

However, the Rarita-Schwinger theory of fields is characterized by a number of pathologies \cite{Velo:1969bt, Adler:2015hna, Adler:2017shl, Adler:2019zxx}, in particular, the singular Dirac bracket turns out to be in the weak-field limit and there is the discontinuity in the number of degrees of freedom when an external field is present. These problems were overcome in \cite{Adler:2017shl} by introducing a field with spin 1/2, which ultimately made it possible to construct a consistent quantum field perturbation theory and calculate the chiral quantum anomaly. An interesting observation is that the coefficient in the chiral anomaly turned out to be 5, which is different from the previous calculations for spin 3/2.

The question that interests us in this work is the manifestation of quantum anomalies in hydrodynamics. In particular, it was shown in a number of works that the chiral vortical effect (CVE) is directly related to the chiral quantum anomaly \cite{Son:2009tf, Sadofyev:2010is, Landsteiner:2011cp, Prokhorov:2020okl, Prokhorov:2020npf, Ambrus:2019khr, Mitkin:2021dxb}. Namely, the coefficient $ \mu^2 $ in the mean value of the axial current in a medium with vorticity corresponds to the coefficient in the chiral anomaly
\begin{eqnarray}
CVE:\quad \langle \hat{j}^{\nu}_A\rangle &=& (A T^2 + C \mu^2) \omega^{\nu}\,,\nonumber \\
Anomaly:\, \langle \partial_{\mu}\hat{j}^{\mu}_A\rangle &=& -\frac{C}{8} \varepsilon^{\mu\nu\alpha\beta}F_{\mu\nu}F_{\alpha\beta}\,.
\label{corresp}
\end{eqnarray}
where $ \omega^{\mu}=\frac{1}{2}\varepsilon^{\mu\nu\alpha\beta}
u_{\nu}\partial_{\alpha}u_{\beta} $ is the vorticity,  $ u_{\mu} $ is the 4-velocity of the fluid, $ \mu $ is the chemical potential and $ T $ is the temperature. This relationship has been well studied in the case of spin 1/2, for which
\begin{eqnarray}
CVE:\quad \langle \hat{j}^{\nu}_A\rangle &=& \left(\frac{1}{6} T^2 + \frac{1}{2\pi^2} \mu^2\right) \omega^{\nu}\,,\nonumber \\
Anomaly:\, \langle \partial_{\mu}\hat{j}^{\mu}_A\rangle &=& -\frac{1}{16\pi^2} \varepsilon^{\mu\nu\alpha\beta}F_{\mu\nu}F_{\alpha\beta}\,.
\label{12}
\end{eqnarray}
Recently a test of the connection with the anomaly was carried out for spin 3/2 for another phenomenon in an external magnetic field, the chiral separation effect (CSE) \cite{Khaidukov:2020mrn}. At the same time, there is no direct verification in the case of higher spins in the case of CVE (\ref{corresp}). 

In this paper, we calculated the CVE for spin 3/2 in Adler's model \cite{Adler:2017shl}, and showed that the coefficient in CVE exactly satisfies the chiral anomaly found in \cite{Adler:2017shl}. In this case, the coefficient 5 is achieved as a combination of the contribution of -1 from the Rarita-Schwinger field itself, and the contribution 6 from the terms of the interaction with the field with spin 1/2. Thus, in the case of CVE, the anomaly is reproduced, but in a qualitatively different way, since the diagrams with a spin 1/2 field did not contribute in \cite{Adler:2017shl, Khaidukov:2020mrn}.

We have shown that in the case of the extended theory of Rarita-Schwinger fields, both the coefficients $ A $ and $ C $ in the formula (\ref{corresp}) differ by a factor of 5 in comparison with the case of spin 1/2 (\ref{12}). This may serve as a sign of a new universality $ A\sim C $ between the temperature-related and the chemical potential-related CVE. This observation differs from the previous predictions \cite{Huang:2018aly}, where linear spin dependence was obtained, since a naive combination of the contributions of fields with spins 3/2 and 1/2, proportional to the spin, results in a factor of 4.

The structure of the work is as follows: in Section II, we present the Lagrangian and currents for the extended theory of spin 3/2 and calculate the corresponding stress-energy tensor, as well as propagators at finite temperature and finite chemical potential. In Section III, using the quantum-statistical density operator for a medium with vorticity, we calculate CVE and show that it corresponds to a chiral anomaly. Section IV discusses the new universality between $ T^2 $ and $ \mu^2 $, the relationship between quantum anomalies and hydrodynamics and contains speculations about the relationship between different types of quantum anomalies. The Conclusion lists the main results.

We use the notations $g_{\mu\nu} = \mathrm{diag} (1, -1, -1, -1)$, $\epsilon^{0123}=1$, in the rest frame $ u_{\mu}=(1,0,0,0) $, and we use the system of units $ e=\hbar=c=k_B=1 $.

\section{The theory of Rarita-Schwinger field coupled to a field with spin 1/2}
\label{sec:theory}

In this section, we present the main relations for the theory of spin 3/2 field, interacting with spin 1/2 field \cite{Adler:2017shl}  (see also \cite{Adler:2019zxx}). The action has the form
\begin{eqnarray}
S=\int d^4 x\,\big( -\varepsilon^{\lambda \rho \mu \nu} \bar{\psi}_{\lambda}\gamma_5 \gamma_{\mu} \partial_{\nu} \psi_{\rho}+i \bar{\lambda} \gamma^{\mu}\partial_{\mu} \lambda - i m \bar{\lambda} \gamma^{\mu}\psi_{\mu}+i m \bar{\psi}_{\mu} \gamma^{\mu}\lambda \big)\,,
\label{action}
\end{eqnarray}
where $ \psi_{\mu} $ is the Rarita-Schwinger field, $ \lambda $ is the field with spin 1/2, $ m $ is the interaction constant. The last two terms describe the interaction of the fields $ \psi_{\mu} $ and $ \lambda $ with the same-sign chirality, which does not allow us to consider them as usual massive terms, and in \cite{Adler:2017shl} they are called ``proto-massive'' terms.

To calculate the stress-energy tensor, it is necessary to go to a curved space-time with an arbitrary metric $ g_{\mu\nu} $ and vary the action with respect to the metric
\begin{eqnarray}
T^{\mu\nu}=-\frac{2}{\sqrt{-g}}\frac{\delta S}{\delta g_{\mu\nu}} \,.
\label{temvar}
\end{eqnarray}
As a result, we obtain the following expression for the symmetric stress-energy tensor, from which we have excluded some terms equal to zero when taking into account the equations of motion  \cite{Das:1978tm, Adler:2015hna}
\begin{eqnarray}
T^{\mu\nu}&=& \frac{1}{2} \varepsilon^{\lambda\alpha\beta\rho}\bar{\psi}_{\lambda}\gamma_5 (\gamma^{\mu}\delta^{\nu}_{\alpha}+\gamma^{\nu}\delta^{\mu}_{\alpha})\partial_{\beta}\psi_{\rho}\nonumber \\
&&+\frac{1}{8}\partial_{\eta}\Big(
\varepsilon^{\lambda\alpha\beta\rho}\bar{\psi}_{\lambda}\gamma_5\gamma_{\alpha} ([\gamma^{\eta},\gamma^{\mu}]\delta^{\nu}_{\beta}+
[\gamma^{\eta},\gamma^{\nu}]\delta^{\mu}_{\beta})\psi_{\rho}\Big)\nonumber \\
&&+\frac{i}{4}\Big(
\bar{\lambda}\gamma^{\nu}\partial^{\mu}\lambda-\partial^{\mu}\bar{\lambda}\gamma^{\nu}\lambda+\bar{\lambda}\gamma^{\mu}\partial^{\nu}\lambda-\partial^{\nu}\bar{\lambda}\gamma^{\mu}\lambda
\Big) \nonumber \\
&&+\frac{i }{2}m\Big(
\bar{\psi}^{\mu}\gamma^{\nu}\lambda-\bar{\lambda}\gamma^{\mu}\psi^{\nu}+\bar{\psi}^{\nu}\gamma^{\mu}\lambda-\bar{\lambda}\gamma^{\nu}\psi^{\mu}
\Big)\,.
\label{temres}
\end{eqnarray}
Currents can be constructed from Noether's theorem and global symmetries $ \psi_{\mu}\to e^{i \alpha+i \beta \gamma_5}\psi_{\mu}$ and $\lambda\to e^{i \alpha+i \beta \gamma_5}\lambda$
\begin{eqnarray}
j^{\mu}&=& i \varepsilon^{\lambda\rho\nu\mu}\bar{\psi}_{\lambda}\gamma_5 \gamma_{\nu}\psi_{\rho} +\bar{\lambda}\gamma_{\mu} \lambda\,, \nonumber \\
j^{\mu}_A&=& -i \varepsilon^{\lambda\rho\nu\mu}\bar{\psi}_{\lambda}\gamma_{\nu}\psi_{\rho} +\bar{\lambda}\gamma_{\mu}\gamma_5 \lambda\,.
\label{cur}
\end{eqnarray}
It is easy to check that the currents and the stress-energy tensor are conserved when the equations of motion are taken into account
\begin{eqnarray}
\partial_{\mu}T^{\mu\nu}=0\,,\,
\partial_{\mu}j^{\mu}=0\,,\,
\partial_{\mu}j^{\mu}_A=0\,.
\label{conser}
\end{eqnarray}
In quantum field theory, the conservation of the axial current is violated by the chiral quantum anomaly, calculated in \cite{Adler:2017shl} in the limit $ m\to \infty $
\begin{eqnarray}
\partial_{\mu}j^{\mu}_A=-\frac{5}{16\pi^2}\varepsilon^{\mu\nu\alpha\beta}F_{\mu\nu}F_{\alpha\beta}\,.
\label{anom}
\end{eqnarray}
Note that the coefficient 5  recently found in \cite{Adler:2017shl} is different from the previous evaluations, and is associated with the interaction with the field $ \lambda $. However, one can see, that it is equal to the sum of the previously known contributions of nonghost part of free spin 3/2 and free spin 1/2 fields.

Propagators at finite temperature can be constructed on the basis of the path integral formalism, developed in \cite{Adler:2017shl}, according to the standard procedure \cite{Laine:2016hma}. When passing to the finite temperature $ T=|\beta|^{-1} $, it is convenient to introduce new notations
\begin{eqnarray}
&&t=-i\tau\,,\quad \mathcal{L}_E(\tau)=-\mathcal{L}_M(t=-i\tau)\,,\quad
\gamma_{\mu}=i^{\delta_{0\mu}-1}\tilde{\gamma}_{\mu}\,,\quad\{\tilde{\gamma}_{\mu}
\tilde{\gamma}_{\nu}\}=2\delta_{\mu\nu}\,,\nonumber \\
&& \tilde{\gamma}_5=\gamma_5=i\gamma^0\gamma^1\gamma^2\gamma^3\,,\quad {\partial}_{\mu}=i^{\delta_{0\mu}}\tilde{\partial}_{\mu}\,,\quad \psi_{\mu}=i^{\delta_{0\mu}}\tilde{\psi}_{\mu}\,,
\nonumber \\
&&P^\pm_{\mu}=(p^{\pm}_n,-{\bf p})\, , \quad p^{\pm}_n=\pi(2n+1)/|\beta|\pm i\mu  \quad (n=0,\pm 1,\pm 2, \cdots) \,, \nonumber \\
&&X_{\mu}=(\tau,-{\bf x})\, , \quad \SumInt_{P}=\frac{1}{|\beta|}\sum_{n=-\infty}^{\infty}\int\frac{d^3p}{(2\pi)^3}\,,\quad  \slashed{P}=P_{\mu}\tilde{\gamma}_{\mu}\,,\quad (P^{+})^2=P^{+}_{\mu}P^{+}_{\mu}\,.
\label{notat}
\end{eqnarray}
The propagators at finite temperature can be found from the path integral written in terms of the Euclidean action and have a form similar to the real-time form in \cite{Adler:2017shl}
\begin{eqnarray}
\langle T_{\tau}\tilde{\psi}_{a \mu}(X_1)\tilde{\bar{\psi}}_{b\nu}(X_2)\rangle_{T} &=& \SumInt_{P}e^{iP^+_{\alpha}(X_1-X_2)^{\alpha}}\frac{i}{2 (P^{+})^2}\Big(\tilde{\gamma}_{\nu} \slashed{P}^+ \tilde{\gamma}_{\mu} +2\Big[\frac{1}{m^2}-\frac{2}{(P^{+})^2}\Big]
P^+_{\mu}P^+_{\nu}\slashed{P}^+\Big)_{ab}\,,\nonumber \\
\langle T_{\tau}\lambda_a(X_1)\tilde{\bar{\psi}}_{b\mu}(X_2)\rangle_{T} &=& \SumInt_{P}e^{iP^+_{\alpha}(X_1-X_2)^{\alpha}}\frac{P^+_{\mu}\slashed{P}^+_{ab}}{m  (P^{+})^2 }\,,\nonumber \\
\langle T_{\tau}\tilde{\psi}_{a\mu}(X_1)\bar{\lambda}_b(X_2)\rangle_{T} &=& \SumInt_{P}e^{iP^+_{\alpha}(X_1-X_2)^{\alpha}}\frac{-P^+_{\mu}\slashed{P}^+_{ab}}{m  (P^{+})^2 }\,,\nonumber \\
\langle T_{\tau}\lambda_a(X_1)\bar{\lambda}_b(X_2)\rangle_{T} &=& 0\,.
\label{prop}
\end{eqnarray}
where $ \mu,\nu $ are Lorentz indices and $ a,b $ are bispinor indices. When deriving (\ref{prop}), we assumed that the subsystems of the fields with spin 3/2 and 1/2 are in equilibrium and $ \mu_{\psi}=\mu_{\lambda}=\mu $. Finally, we note that following \cite{Adler:2017shl, Adler:2019zxx}, the ghost fields should be considered non-propagating and non-interacting with the rest of the fields, due to which ghosts do not contribute to the quantities we are considering.

\section{Chiral vortical effect for spin 3/2}
\label{sec:calc}

In this section, we calculate CVE for spin 3/2 field interacting with spin 1/2 field. The properties of the medium in the state of global thermodynamic equilibrium are described by the density operator of Zubarev \cite{Zubarev:1979, Buzzegoli:2017cqy, Prokhorov:2019cik, Buzzegoli:2020fjm}
\begin{eqnarray}
&&\hat{\rho}=\frac{1}{Z}\exp\Big\{-\beta_{\mu}(x)\hat{P}^{\mu}
+\frac{1}{2}\varpi_{\mu\nu}\hat{J}^{\mu\nu}_x+\zeta  \hat{Q}
\Big\} \,,
\label{rho}
\end{eqnarray}
where $ \varpi_{\mu\nu}=-\frac{1}{2}(\partial_{\mu}\beta_{\nu}-\partial_{\nu}\beta_{\mu}) $ is the thermal vorticity tensor, $ \zeta=\frac{\mu}{T} $, $\hat{P}^{\mu}$ is the 4-momentum operator, $\hat{Q}$ is the charge operator, and $\hat{J}^{\mu\nu}_x$ are the Lorentz transformation generators shifted by the vector $x^{\mu}$, which are expressed in terms of the shifted operators of the stress-energy tensor
\begin{eqnarray}
&& \hat{J}^{\mu\nu}_x=
\int d\Sigma_{\lambda}\big[y^{\mu}\hat{T}^{\lambda\nu}_x(y)-y^{\nu}\hat{T}^{\lambda\mu}_x(y)\big] \,.
\label{J}
\end{eqnarray}
where $ d\Sigma_{\lambda} $ is a volume element on an arbitrary spacelike hypersurface (arbitrariness of the hypersurface follows from the conditions of global thermodynamic equilibrium). Density operator (\ref{rho}) provides a universal and fundamental approach to the description of effects in a relativistic moving and charged medium \cite{Prokhorov:2019yft, Becattini:2021iol, Prokhorov:2019cik, Prokhorov:2018bql, Becattini:2019dxo, Buzzegoli:2020fjm, Palermo:2021hlf}. In particular, it was used to find a lot of chiral effects \cite{Buzzegoli:2017cqy, Buzzegoli:2018wpy}, corrections to them \cite{Prokhorov:2018bql}, and also to prove the Unruh effect from the point of view of statistics as well as the duality between statistics and field theory in a space with a conical singularity \cite{Prokhorov:2019cik, Prokhorov:2019yft, Prokhorov:2019hif, Becattini:2017ljh}. 


Operator (\ref{rho}) can be used to study the effects of acceleration and vorticity, since
\begin{eqnarray}
\varpi_{\mu\nu}\hat{J}^{\mu\nu} = \frac{1}{T}(-2 a^{\rho} \hat{K}_{\rho}-2 \omega^{\rho} \hat{J}_{\rho})\,,
\label{wJ1}
\end{eqnarray}
where $ a^{\rho} $ -- acceleration, $ \omega^{\rho} $ -- vorticity, $  \hat{K}_{\rho} $ -- boost operator, and $ \hat{J}_{\rho} $ -- operator of angular momentum. We will be interested in the effects of vorticity and angular momentum
\begin{eqnarray}
\hat{J}^{\mu}=-\frac{1}{2} \epsilon^{\mu\nu\rho\sigma}u_{\nu}\hat{J}_{\rho\sigma}\,. 
\label{Jdec}
\end{eqnarray}
In a particular case of rigidly rotating medium, the operator (\ref{rho}) can be transformed to the more well-known form of the density operator for an equilibrium rotating medium \cite{Vilenkin:1979ui, Vilenkin:1980zv, Ambrus:2021eod}, but (\ref{rho}) gives a more general relativistic form of density operator.
For the axial current $ \hat{j}_A^{\mu} $ in the first order of the perturbation theory formulas (\ref{rho})-(\ref{Jdec}) give (see \cite{Buzzegoli:2017cqy, Prokhorov:2018bql} for details)
\begin{eqnarray} \label{j5pertall}
\langle\hat{j}_A^{\mu}\rangle^{(1)} &=& W\, \omega^{\mu}\,,\nonumber \\  
W  &=&  \int_0^{|\beta|} d\tau \langle T_{\tau} \hat{J}^{3}_{-i\tau u}\hat{j}_A^{3}(0)\rangle_{T,c}=  C^{023|1}-C^{013|2}\,,\nonumber \\ 
C^{\alpha\beta\gamma|i} &=& \int_0^{|\beta|} d\tau \int d^3x\, x^i \langle T_{\tau} \hat{T}^{\alpha \beta}(-i\tau,{\bf x}) \hat{j}_A^{\gamma}(0)\rangle_{T,c} \,,
\label{first}
\end{eqnarray}
where the scalar coefficient $ W $ can be evaluated in the rest frame $ \beta_{\mu}=(T^{-1},0,0,0) $, which is expressed by the subscript $ T $, and in the following we denote $ \hat{T}^{\alpha \beta}(-i\tau,{\bf x})\to \hat{T}^{\alpha \beta}(\tau,{\bf x}) $. Now the main goal is to find the correlators of the form $ C^{\alpha\beta\gamma|i} $. To do this, let's first split $ \hat{T}^{\mu\nu} $ and $ \hat{j}_A^{\mu} $ into terms with a different set of fields
\begin{eqnarray}
\hat{T}^{\mu\nu}& = & \hat{T}^{\mu\nu}_{\bar{\psi}\psi}+\hat{T}^{\mu\nu}_{\bar{\lambda}\lambda}+\hat{T}^{\mu\nu}_{\bar{\psi}\lambda}+\hat{T}^{\mu\nu}_{\bar{\lambda}\psi}\,, \nonumber \\
\hat{j}_A^{\mu} &=& \hat{j}_{A\bar{\psi}\psi}^{\mu}+ \hat{j}_{A\bar{\lambda}\lambda}^{\mu}\,,
\label{temparts}
\end{eqnarray}
where the notation is obvious. Then we get that $ W $ is split into 8 terms depending on the set of the fields
\begin{eqnarray}
W&=&  W_{\bar{\psi}\psi\bar{\psi}\psi}+W_{\bar{\psi}\psi\bar{\lambda}\lambda}+W_{\bar{\lambda}\lambda\bar{\psi}\psi}+W_{\bar{\lambda}\lambda\bar{\lambda}\lambda}+W_{\bar{\psi}\lambda\bar{\psi}\psi}+W_{\bar{\psi}\lambda\bar{\lambda}\lambda}+W_{\bar{\lambda}\psi\bar{\psi}\psi}+W_{\bar{\lambda}\psi\bar{\lambda}\lambda}\,,
\label{wparts}
\end{eqnarray}
where the first two indices denote fields in $ \hat{T}^{\mu\nu} $, and the second two -- in $ \hat{j}_A^{\mu} $. From the equality $ \langle \lambda \bar{\lambda}\rangle = 0 $ it is obvious that $ W_{\bar{\lambda}\lambda\bar{\lambda}\lambda}=W_{\bar{\lambda}\psi\bar{\lambda}\lambda}=W_{\bar{\psi}\lambda\bar{\lambda}\lambda}= 0$. Since we are interested in the limit of $ m\to \infty $, it is also clear in advance that $ W_{\bar{\psi}\psi\bar{\lambda}\lambda},W_{\bar{\lambda}\lambda\bar{\psi}\psi}\to 0$ at $ m \to \infty $. Thus, only three terms remain $W = W_{\bar{\psi}\psi\bar{\psi}\psi}+W_{\bar{\psi}\lambda\bar{\psi}\psi}+W_{\bar{\lambda}\psi\bar{\psi}\psi}$ .

Let's describe the main steps of the following derivation for $ W_{\bar{\psi}\psi\bar{\psi}\psi} $. All operators are to be presented in split form
\begin{eqnarray}\label{Tsplit}  \nonumber 
\hat{T}^{\sigma\tau}_{\bar{\psi}\psi}(X) &=& \lim_{\scriptscriptstyle X_1,X_2\to X} i^{\delta_{0\eta}+\delta_{0\xi}} \mathcal{D}^{\sigma\tau\eta\xi}_{(\bar{\psi}\psi)a b}(\partial_{X_1},\partial_{X_2})\tilde{\bar{\psi}}_{\eta a}(X_1)\tilde{\psi}_{\xi b}(X_2)\,,\\ 
\mathcal{D}^{\sigma\tau\eta\xi}_{(\bar{\psi}\psi)}(\partial_{X_1},\partial_{X_2})&=& \frac{1}{2} i^{1-\delta_{0\sigma}+\delta_{0\beta}}\varepsilon^{\eta \xi \tau\beta}\Big(\gamma_5 \tilde{\gamma}_{\sigma} \tilde{\partial}_{\beta}^{X_2}-\frac{1}{4}\gamma_5 \tilde{\gamma}_{\beta}[\tilde{\gamma}_{\vartheta},\tilde{\gamma}_{\sigma}] (\tilde{\partial}^{X_1}_{\vartheta}+\tilde{\partial}^{X_2}_{\vartheta})\Big)+(\sigma\leftrightarrow \tau)\,, \nonumber \\
\hat{j}^{\sigma}_{A\bar{\psi}\psi}(X) &=& \lim_{\scriptscriptstyle X_1,X_2\to X} i^{\delta_{0\eta}+\delta_{0\xi}} \mathcal{J}^{\sigma \eta \xi}_{A(\bar{\psi}\psi)a b}\tilde{\bar{\psi}}_{\eta a}(X_1)\tilde{\psi}_{\xi b}(X_2)\,,\nonumber \\
\mathcal{J}^{\sigma \eta \xi}_{A(\bar{\psi}\psi)a b}&=&-i^{\delta_{0\mu}}\varepsilon^{\eta \xi \mu \sigma}
\tilde{\gamma}_{\mu}\,.
\label{split}
\end{eqnarray}
Substituting (\ref{split}) into (\ref{first}), we get
\begin{eqnarray}
C^{\alpha\beta\gamma|i}_{(\bar{\psi}\psi)} &=& \int_0^{|\beta|} d\tau \int d^3x\, x^i 
\lim_{
	\def\arraystretch{0.5}\begin{array}{ll}
	{\scriptscriptstyle X_1,X_2\to X}\vspace{0.1mm}\\
	{\scriptscriptstyle Y_1,Y_2\to 0}
	\end{array}} i^{\delta_{0\eta}+\delta_{0\xi}+\delta_{0\rho}+\delta_{0\lambda}}
\mathcal{D}^{\alpha\beta\eta\xi}_{(\bar{\psi}\psi)a b}(\partial_{X_1},\partial_{X_2})\mathcal{J}^{\gamma \rho \lambda}_{A(\bar{\psi}\psi)cd}\times  \nonumber \\
&& \times \langle T_{\tau} \tilde{\bar{\psi}}_{\eta a}(X_1)\tilde{\psi}_{\xi b}(X_2)\tilde{\bar{\psi}}_{\rho c}(Y_1)\tilde{\psi}_{\lambda d}(Y_2) \rangle_{T,c} \,.
\label{c}
\end{eqnarray}
Using the propagators (\ref{prop}), we obtain
\begin{eqnarray}
C^{\alpha\beta\gamma|i}_{(\bar{\psi}\psi)} &=& \int_0^{|\beta|} d\tau \int \frac{d^3xd^3pd^3q}{(2\pi)^6}\, x^i 
\frac{1}{|\beta|^2}\sum_{\def\arraystretch{0.5}\begin{array}{ll}
	{\scriptscriptstyle p_n=\pi(2n+1)}\vspace{0.1mm}\\
	{\scriptscriptstyle q_l=\pi(2l+1)}
	\end{array}}
\frac{-1}{4 (P^{+})^2(Q^{-})^2} 
i^{\delta_{0\eta}+\delta_{0\xi}+\delta_{0\rho}+\delta_{0\lambda}}
e^{i(p_n  + q_l )\tau}\times
\nonumber \\
&&\times e^{-i(\bold{p}+\bold{q})\bold{x}}\mathrm{tr}\Big[
\mathcal{D}^{\alpha\beta\eta\xi}_{(\bar{\psi}\psi)}(iQ^{-},i P^{+})
\Big\{\tilde{\gamma}_{\rho} \slashed{P}^{+}\tilde{\gamma}_{\xi} +2\Big[\frac{1}{m^2}-\frac{2}{(P^{+})^2}\Big]
P^+_{\rho}P^+_{\xi}\slashed{P}^{+}\Big\}\times
\nonumber \\
&&\times \mathcal{J}^{\gamma \rho \lambda}_{A(\bar{\psi}\psi)}
\Big\{\tilde{\gamma}_{\eta} \slashed{Q}^{-}\tilde{\gamma}_{\lambda} +2\Big[\frac{1}{m^2}-\frac{2}{(Q^{-})^2}\Big]
Q^-_{\eta}Q^-_{\lambda}\slashed{Q}^{-}\Big\}\Big]\,.
\label{crez}
\end{eqnarray}
Summation over the Matsubara frequencies should be made taking into account the poles $ ((p_n\pm i\mu)^2+E^2)^{-r} $, where $ r=1,2 $ according to the formulas from Appendix A.4 of \cite{Buzzegoli:2020fjm} \footnote{GP thanks M. Buzzegoli for discussing this issue.}, and the explicit dependence on the coordinate $ x^i $ can be absorbed into the derivative of the exponent $ \frac{\partial}{\partial p^{i}} e^{-i\bold{p}\bold{x}} $. After that, the integration and summation over one of the momenta is removed by the delta function. Finally, integration over the angles at $ d^3p = \sin( \vartheta) p^2 \, dp \,d\phi\,d\vartheta $ and integration over $ \tau $ can be done directly.

As a result, we get that each of the coefficients is expressed as a combination of an infinite and a finite integral over the momentum
\begin{eqnarray}
W_{\bar{\psi}\psi\bar{\psi}\psi} &=&  -\frac{2 }{3 \pi^2}\int_0^{\infty} p\, dp
+\int_0^{\infty} \frac{dp}{\pi^2 T}\Bigg( 
-\frac{2p^3}{3T}\big[n_{F}(x)^{3}+n_{F}(y)^{3}\big]
+\Big(\frac{p^3}{ T}-\frac{p^2}{6}\Big)\big[n_{F}(x)^{2} \\
&&+n_{F}(y)^{2}\big] +\Big(-\frac{p^3}{3T}+\frac{p^2}{6}+\frac{2 p T}{3}\Big)\big[n_{F}(x)+n_{F}(y)\big]
\Bigg)\,, \nonumber \\
W_{\bar{\psi}\lambda\bar{\psi}\psi} &=&  -\frac{1}{3 \pi^2}\int_0^{\infty} p\, dp
+\int_0^{\infty} \frac{dp}{\pi^2 T}\Bigg( 
\frac{2p^3}{3T}\big[n_{F}(x)^{3}+n_{F}(y)^{3}\big]
+\Big(-\frac{p^3}{ T}-\frac{p^2}{3}\Big)\big[n_{F}(x)^{2}\nonumber \\
&&+n_{F}(y)^{2}\big] +\Big(\frac{p^3}{3T}+\frac{p^2}{3}+\frac{p T}{3}\Big)\big[n_{F}(x)+n_{F}(y)\big]
\Bigg)\,, \nonumber \\
W_{\bar{\lambda}\psi\bar{\psi}\psi} &=&  \frac{1}{ \pi^2}\int_0^{\infty} p\, dp
+\int_0^{\infty} \frac{dp}{\pi^2 T}\Big(
- 2p^2\big[n_{F}(x)^{2}+n_{F}(y)^{2}\big]
+(2 p^2 - p T )\big[n_{F}(x)+n_{F}(y)\big]
\Big)\,,  \nonumber
\label{wres}
\end{eqnarray}
where $ n_{F}(E)=(1+e^{E/T})^{-1} $ is the Fermi-Dirac distribution, $ x=p+\mu,\, y=p-\mu$. The finite parts can be found analytically as they are expressed in terms of polynomial combinations of polylogarithms \cite{Prokhorov:2019hif, Stone:2018zel}. As a result, we get
\begin{eqnarray}
W_{\bar{\psi}\psi\bar{\psi}\psi} &=&  -\frac{2}{3 \pi^2}\int_0^{\infty} p\, dp
-\frac{T^2}{6}-\frac{\mu^2}{2 \pi^2}\,, \nonumber \\
W_{\bar{\psi}\lambda\bar{\psi}\psi} &=&  -\frac{1}{3 \pi^2}\int_0^{\infty} p\, dp
+\frac{T^2}{2}+\frac{3 \mu^2}{2 \pi^2}\,, \nonumber \\
W_{\bar{\lambda}\psi\bar{\psi}\psi} &=&  \frac{1}{ \pi^2}\int_0^{\infty} p\, dp
+\frac{T^2}{2}+\frac{3 \mu^2}{2 \pi^2}\,.
\label{wres1}
\end{eqnarray}
Despite the fact that each of the terms has an ultraviolet divergence, the sum is finite. Thus, ultraviolet divergences appear at intermediate stages of calculations, but mutually cancel out between different contributions in the final formula for the physical effect. Also the integral itself is greatly simplified
\begin{eqnarray}
W_{\bar{\psi}\psi\bar{\psi}\psi}+
W_{\bar{\psi}\lambda\bar{\psi}\psi} +
W_{\bar{\lambda}\psi\bar{\psi}\psi} = -\frac{5}{2 \pi^2}\int_0^{\infty} p^2 dp \Big[n_{F}(x)^{\prime}+n_{F}(y)^{\prime}\Big] = \frac{5 T^2}{6}+  \frac{5 \mu^2}{2\pi^2}\,,
\label{wsum}
\end{eqnarray}
where $ n_F(E)^{\prime}=\frac{d}{dE} n_F(E)$. As a result, we obtain the following expression for the axial current, which corresponds exactly to the chiral anomaly (\ref{anom})
\begin{eqnarray}
\langle \hat{j}^{\nu}_A\rangle^{(1)} = \left( \frac{5 T^2}{6}+  \frac{5 \mu^2}{2\pi^2}\right) \omega^{\nu}\,.
\label{jres}
\end{eqnarray}
In this case, the coefficients 5 in terms $ T^2 $ and $ \mu^2 $ were obtained as a result of summation 6-1=5, where 6 is the contribution of the interaction terms, and -1 is the contribution of the pure Rarita-Schwinger field. This distinguishes the above calculation from the calculation of the chiral anomaly and CSE in \cite{Adler:2017shl, Khaidukov:2020mrn}, where the additional field did not contribute. Thus, the anomaly is reconstructed, but in a qualitatively different way.

\section{Discussion: quantum anomalies in hydrodynamics}
\label{sec: disc}

We have shown an exact correspondence between hydrodynamics and quantum field theory: the coefficient in front of the chiral anomaly (\ref{anom}) corresponds to the coefficient in CVE (\ref{jres}). Such a correspondence of the two theories is not accidental and was predicted in a number of papers \cite{Son:2009tf, Sadofyev:2010is}. Our result confirms the accuracy of the predictions made not only in the case of spin 1/2, but also for higher spins and demonstrates how this correspondence is realized at the level of microscopic theory.

In particular, it was shown in \cite{Sadofyev:2010is} that hydrodynamics can be considered as an effective field theory with additional interaction corresponding to the substitution
\begin{eqnarray}
e A_{\nu}\to e A_{\nu}+\mu \cdot u_{\nu}\,.
\label{subs}
\end{eqnarray}
Using the well-known expression for the chiral anomaly, but now for the effective field (\ref{subs}), one can clearly obtain a number of chiral phenomena and demonstrate their  relation to the anomaly (\ref{corresp}).

In a different context, the same relationship was substantiated from the point of view of the equations of relativistic hydrodynamics and the second law of thermodynamics in \cite{Son:2009tf}. Equation (\ref{anom}) is to be included into the system of equations of hydrodynamics. From the condition of non-negativity of the divergence of the entropy current it follows that the currents arise, directly related to the anomaly.

Both \cite{Son:2009tf, Sadofyev:2010is} approaches deal with the $ \mu^2 $ term, while the temperature term $ T^2 $ and the coefficient $ A $  in (\ref{corresp}) are assumed to be associated with either the gravitational anomaly \cite{Stone:2018zel, Golkar:2012kb, Prokhorov:2020okl, Landsteiner:2011cp, Prokhorov:2020npf} or the global one \cite{Golkar:2015oxw}. In particular, \cite{Stone:2018zel} considers radiation from an analogue of a rotating black hole, a quantum anomaly on the horizon of which serves as a pump that creates an anomalous axial current. Such a relationship between the $T^2$ term and the gravitational anomaly has been verified for the case of spin 1/2  \cite{Stone:2018zel}, and recently for spin 1 \cite{Prokhorov:2020npf}.

So, on the one hand, the relationship between the coefficient $ A $ and the gauge chiral anomaly and $ C $ with the gravitational chiral anomaly, allows us to make a hypothesis that the gravitational chiral anomaly is also 5 times higher for system of interacting spin 3/2 and spin 1/2 fields than for spin 1/2. An indication of the possible existence of such a relationship between gravitational and gauge anomalies was recently found in a completely different context in \cite{Volovik:2021myq}.

On the other hand, the comparison of (\ref{jres}) and (\ref{12}) indicates the existence of possible universality $ A\sim C $. From the point of view of the chiral kinetic approach \cite{Huang:2018aly}, the proportionality $ A\sim C $ follows from the spin-vorticity effective coupling $\vec{S}\cdot \vec{\Omega}$. However, in this case the coefficients are to be proportional to the spin $ A\sim C\sim S $. Naively for a system of fields with spins $ 3/2 $ and $ 1/2 $, which we consider, this would lead to a factor of 4, not 5. Thus, the universality $ A\sim C $ is probably a more general phenomenon than $ A\sim C\sim S $.


In \cite{Khaidukov:2020mrn} another phenomenon, the CSE, was calculated in the framework of the theory \cite{Adler:2017shl}, and the result was
\begin{eqnarray}
CSE:\quad\langle \hat{j}_A^{\nu}\rangle =  \frac{5 \mu}{2\pi^2} B^{\nu}\,,
\label{cse}
\end{eqnarray}
where $ B^{\mu} $ is the magnetic field. Thus, the CSE also satisfies the chiral anomaly (\ref{anom}). Technically the correspondence between the CSE and the anomaly is clear, since both of them can be described by the same diagrams, but with the replacement of one of the fields in one of the vertices by the chemical potential in the case of CSE. In the case of CVE, similar reasoning cannot be used, since the operators of the stress-energy tensor are located in the vertices instead of the current operators.

The relationship between CSE and anomalies (\ref{anom}) follows from \cite{Son:2009tf, Sadofyev:2010is} and also follows from the recent analysis in \cite{Buzzegoli:2020fjm}. The conditions of global thermodynamic equilibrium fix a direct connection between the chemical potential and the external electromagnetic field
\begin{eqnarray}
\zeta(x)=\zeta_0-\beta_{\sigma}F^{\lambda\sigma}x_{\lambda}+\frac{1}{2}\varpi_{\sigma\rho}x^{\rho}F^{\lambda\sigma}x_{\lambda}\,,
\label{chem}
\end{eqnarray}
where $ \zeta=\frac{\mu}{T} $. Using (\ref{chem}) for differentiating (\ref{cse}) results in exactly (\ref{anom}).

\section{Conclusion}
\label{sec: concl}

In \cite{Adler:2017shl} the extension of the Rarita-Schwinger field theory was developed, in which an interaction with an additional field with spin $ 1/2 $ was introduced to overcome some of the pathologies of the theory and calculate chiral anomaly. We consider this theory at finite temperature, finite chemical potential, and nonzero vorticity. We calculated the chiral vortical effect in this theory and verified that the coefficient in front of the $ \mu^2 $ term corresponds to the coefficient 5 in the chiral anomaly. We have clearly demonstrated that this result is achieved by summing of the contribution of the interaction terms equal to 6, and the contribution from only the Rarita-Schwinger field equal to -1. We hypothesized a gravitational anomaly for a spin $ 3/2 $ field interacting with a spin $ 1/2 $ field in this new framework: if the term $ T^2 $ is associated with a gravitational anomaly, it can be expected to be five times that of spin 1/2.

Comparison of the formulas for CVE for the extended theory of spin 3/2 and in the case of spin 1/2 suggests the existence of a new universality between the coefficients $ A\sim C $ of the vortical effects associated with temperature ${j}_A^{\nu}=\ A\cdot T^2 \omega^{\nu} $ and with chemical potential ${j}_A^{\nu}=C\cdot \mu^2 \omega^{\nu}$.

We have also demonstrated that there is a cancellation of ultraviolet divergences between the contributions to CVE from the term with only the Rarita-Schwinger field and the terms of interaction, each of which diverges separately.

\section{Acknowledgments}
The authors are thankful to A. I. Vainshtein 
and M. Buzzegoli for valuable discussions. The work was supported by Russian Science Foundation Grant No 21-12-00237.

\bibliography{lit}

\end{document}